\documentclass[letterpaper, 10 pt, conference]{ieeeconf}  

\IEEEoverridecommandlockouts                              

\usepackage{cite}
\usepackage{amsmath,amssymb,amsfonts}
\usepackage{algorithmic}
\usepackage{graphicx}
\usepackage{textcomp}
\usepackage{xcolor}
\def\BibTeX{{\rm B\kern-.05em{\sc i\kern-.025em b}\kern-.08em
    T\kern-.1667em\lower.7ex\hbox{E}\kern-.125emX}}
\begin{document}

\title{\LARGE \bf
Segmentation method for cerebral blood vessels from MRA using hysteresis}

\author{Georgia Kenyon$^{1,3}$, Stephan Lau$^{1,2}$, Michael A. Chappell$^{3}$ and Mark Jenkinson$^{1,2}$
\thanks{$^{1}$Australian Institute for Machine Learning, School of Computer and Mathematical Sciences, The University of Adelaide, Australia, 
        {\tt\small georgia.kenyon@adelaide.edu.au}}%
\thanks{$^{2}$South Australian Health and Medical Research Institute, Adelaide, Australia}
\thanks{$^{3}$Sir Peter Mansfield Imaging Centre, School of Medicine, University of Nottingham, UK}%
}


\maketitle
\thispagestyle{empty}
\pagestyle{empty}

\begin{abstract}
Segmentation of cerebral blood vessels from Magnetic Resonance Imaging (MRI) is an open problem that could be solved with deep learning (DL). However, annotated data for training is often scarce. Due to the absence of open-source tools, we aim to develop a classical segmentation method that generates vessel ground truth from Magnetic Resonance Angiography for DL training of segmentation across a variety of modalities. The method combines size-specific Hessian filters, hysteresis thresholding and connected component correction. The optimal choice of processing steps was evaluated with a blinded scoring by a clinician using 24 3D images. The results show that all method steps are necessary to produce the highest (14.2/15) vessel segmentation quality score. Omitting the connected component correction caused the largest quality loss. The method, which is available on GitHub, can be used to train DL models for vessel segmentation.  


\indent \textit{Clinical relevance}— We provide and validate an open-access method to efficiently segment vessels from brain MRA images for improved diagnostics and intervention planning. 


\end{abstract}

\section{INTRODUCTION}

Cerebrovascular diseases such as aneurysms and stenosis affect blood flow and vessels in the brain. This may lead to a lack of sufficient blood flow, and as a consequence cause stroke. Optimal treatment methods require imaging techniques such as 3D time-of-flight (TOF) magnetic resonance angiography (MRA) prior to intervention. However, visualization of blood vessels from raw data - without appropriate pre-processing can be difficult, due to noise, uneven contrast and limited signal from smaller vessels. This requires high-level, automated pre-processing and segmentation methods to be developed. 


Segmentation can be a challenging task, particularly for images of lower quality. Deep learning provides a solution to this, as models can learn more abstract patterns, such as vessels presented as small voxel intensity changes, from data directly. However, DL requires large amounts of labelled data for training, which is often infeasible. Manual annotation of vessels in 3D images is a very tedious, time-consuming job that requires a high degree of expertise- with labeling taking 60–80 min per MRA image~\cite{livne2019u}. Due to this, developing (semi)-automatic approaches to provide vascular ground truth to supervised approaches has become an important task. 

Some existing works utilize DL for the segmentation of MRA, applying sparse label techniques~\cite{zhang2020cerebrovascular}, synthetic or limited training sets~\cite{livne2019u} or unsupervised learning~\cite{fan2020unsupervised}, but these have not been made openly available. Existing vessel data sets are limited or unsuitable for semantic segmentation tasks. For example, Brava~\cite{wright2013digital}, an open-source vessel reconstruction data set, contains reconstructions, but without the paired MR images necessary for DL training. The freely available 'IXI' data set~\cite{braindevelopment}, offers a large subset of MRA, along with additional paired MR acquisitions, however these are unprocessed and do not contain voxel-wise segmentations. 

Multiple works have been published that use a combination of traditional segmentation methods on MRA. They apply pre-processing filters that highlight 'vesselness'- tubular and high contrast features, use thresholding techniques and apply voxel correction~\cite{li2020statistical, lu2016vessel}. However, there are no open-source methods easily available for reuse. In 2016, Lu et al.~\cite{lu2016vessel} utilized the Hessian filter, to enhance the vessels in MRA, prior to segmentation. This filter uses edge-based enhancement for tubular structures, by taking advantage of the strong gradients at the edges of vessels in comparison to the background. In 2020, Avadiappan et al.~\cite{avadiappan2020fully} employed an alternative method to implement the concept of 'vesselness', using an adaptive Frangi filter to preserve the radii of vessels. The adaptive filter was implemented to quantify vessel radii, as it was summarized that existing methods underestimated the segmentation for vessels of larger radii and overestimated thin vessels.
Multiple methods apply thresholding techniques to extract vessels, applying Gaussian Mixture models over the intensity distributions to separate anatomic structures~\cite{lu2016vessel, wen2015novel, li2020statistical}. However, these single threshold methods can be limited when extracting fine detail. Thresholding with hysteresis, developed for the Canny Edge detector~\cite{canny1983finding}, applies a high and low threshold, to preserve intensity information. Chang et al.~\cite{chang2008small} applied a modified version of the Canny Edge Detector to detect small retinal blood vessels, generating a dynamic hysteresis threshold value based on local neighborhood of a pixel, improving the detection of smaller edges. In 2021, Ooi et al.~\cite{ooi2021interactive} applied the edge detector to segment retinal images, after hand-selecting the vessels, to highlight finer details and draw out vessel edges. 
Voxel correction methods, such as erosion, are also applied to segmentation outputs to remove noise after thresholding~\cite{lu2016vessel}. However, small vessel misclassification as noise can limit these methods. 

\begin{figure}[tbp]
\centering
\parbox{0.6\columnwidth}{\includegraphics[width=0.6\columnwidth,trim={0 0 0 -6mm}]{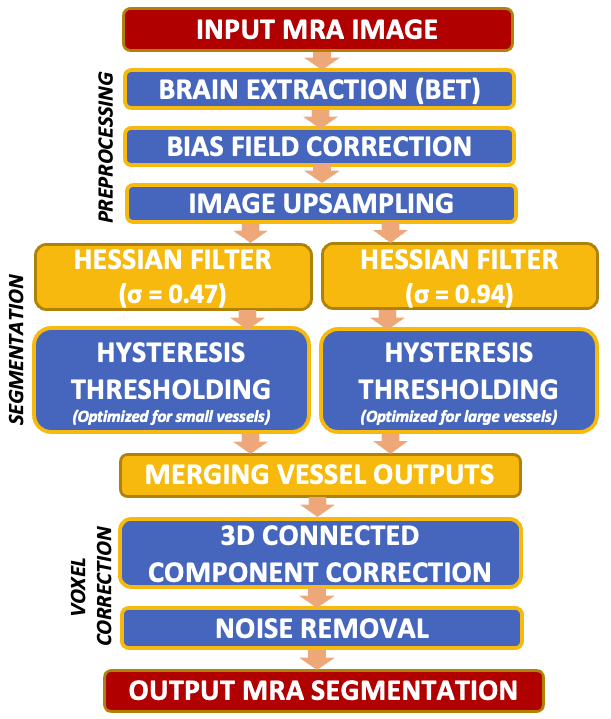} }
\qquad
\begin{minipage}{0.27\columnwidth}%
    \includegraphics[width=\columnwidth]{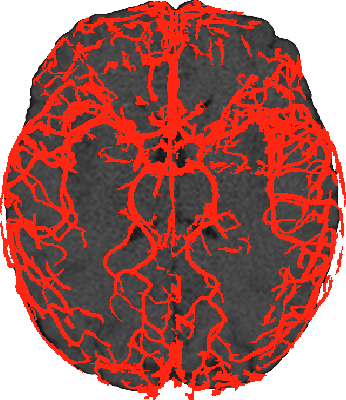}
\end{minipage}%
\caption{Proposed method structure (left) with axial view of cerebral blood vessels (red) segmented with the proposed method from an example MRA (right) shown with one MRA slice (grey) for anatomical reference.}
\label{fig:pipe}
\end{figure}

This paper aims to develop a method, shown in Figure~\ref{fig:pipe}, to segment cerebral blood vessels from MRA images. The method provides a solution to data labelling required to train deep learning models for segmentation tasks using non-MRA modalities (e.g. T2-weighted images) by utilizing paired data sets such as IXI. The method includes a novel combination of processing elements that exploit the benefits of the Hessian Matrix to enhance vessels regardless of vessel radii, and segment using hysteresis thresholds. The method also utilizes the benefits of connected component correction to remove noise and artifacts, whilst maintaining small vessel integrity.

\section{MATERIALS \& METHODS}


\subsection{Imaging Data \& Pre-processing}

3D TOF-MRA was extracted from the IXI database~\cite{braindevelopment}.  Only 3T and 1.5T images acquired in Hammersmith Hospital (IXI-HH) or Guy's Hospital (IXI-Guys), with a resolution of 0.47 x 0.47 x 0.8 $mm^3$, were used. 
The Brain Extraction Tool (BET) in FSL~\cite{JENKINSON2012782} was used to skull-strip the TOF-MRA, to exclude non-vascular skull voxels of similar intensity. Bias field correction was then applied, using FSL's FAST.

\subsection{Vessel-Enhancement Filter}

A 3D line-enhancement filter based on the eigenvalues of the Hessian Matrix was used to enhance the tubular structure of vessels in the image~\cite{sato19973d}. An anisotropic Gaussian filter was applied during Hessian, to reduce noise, and was varied using the algorithm's standard deviation parameter $\sigma$. The $\sigma$ values were chosen based on the image resolution, to optimize the extraction of the vessels specific to their radii~\cite{deshpande2021automatic}, since low $\sigma$ values would highlight small, thin structures, whilst larger values would highlight the larger structures. 
Due to the variation of the vessel structures within the MRA, two values for sigma were applied, producing two image outputs per MRA input. The value $\sigma_{l} = 0.47$ was chosen to conserve fine vascular detail of the small vessels, whilst $\sigma_{u} = 0.94$ was chosen to promote the tubular vessels with larger diameters, reaching two voxels, such as in the Circle of Willis~\cite{sato19973d}. 

\subsection{Hysteresis Thresholding}
Both vessel enhancement outputs were thresholded at a fraction of the 99.9th percentile intensity of the image, respectively. The intensity value at the 99.9th percentile was used instead of the maximum to prevent high intensity noise and outliers impacting this value. The decision to use fractions of the intensity was made so that the threshold applied uniformly across the data set. Importance was placed on improving the quality of the segmentation for larger vessels from the $\sigma_{u}$ filtered images, whilst the extraction of finer vessels was the focus for thresholding the $\sigma_{l}$ filtered images. The absolute value of the threshold is obtained from the relative threshold ${th_{p}}$ by applying it to the image's 99.9th percentile intensity value. 



Hysteresis thresholding was applied to improve the detection of smaller edges from the vessels. The method took two threshold values: a low-threshold value (LTV) and a high-threshold value (HTV). A voxel was included if its intensity was above the HTV, or between the LTV and HTV and connected to a voxel above the HTV.

To preserve the fine detail of the smaller vessels for the $\sigma_{l}$ outputs, $\overline{th_{p}}$ was applied as the HTV (e.g. HTV = 67), whilst the LTV was chosen at 10\% below $\overline{th_{p}}$ (e.g. LTV = 57). The lower range was chosen to promote inclusion of extra voxels of lower intensities if they were connected to an established edge.
The hysteresis threshold range for the data produced using the $\sigma_{u}$ filter, utilized the $\overline{th_{p}}$ as the low-threshold value for hysteresis (e.g. LTV = 39), and a HTV value ten percent above this (e.g. HTV = 49). This was chosen to remove unwanted noise and extract the vessels with larger intensity variations. 
All vessel data was binarized after hysteresis, with background and brain voxels set to zero, and vessels set to one. The two binary outputs were then merged (with a union operation) to form one image. 

\subsection{Connected Component Selection}

A voxel correction algorithm was implemented, to reduce the noise and non-vessel structures left behind after hysteresis thresholding. This was achieved using 3D connected component correction, which exploits the continuous structure of the blood vessels to remove small, disconnected voxel clusters of high intensity left behind by thresholding, suspected to be noise. Any clusters of size less than $ 10mm^3 $ were removed.

\subsection{Clinical Expert Scoring}

To determine an optimal set of hysteresis thresholds, outputs over a range of values were presented to a clinical expert, who then chose the option with the highest anatomical accuracy.

An ablation experiment was also performed, removing the four steps of the method one at a time, and the 24 images from the IXI data set were reprocessed by each alternative method. 3D segmentation outputs were displayed to a trained clinician using FSLeyes~\cite{JENKINSON2012782} and overlaid over the original MRA to enhance detail visibility. The clinician was instructed to score the output images of the alternative methods using metrics determined from literature~\cite{MOCCIA201871}, as shown in Table~\ref{tabScoreDefinition}. 


\begin{table}[bt]
\vspace*{2.5mm} 
\caption{Quality Scores for Ablation Experiment}
\vspace*{-6mm} 
\label{tabScoreDefinition}
\begin{center}
\begin{tabular}{|p{1.5cm}|p{6.5cm}|}
\hline
\textbf{\textit{Metrics}} & \textbf{\textit{Description: score out of 3}}\\
\hline
Connectivity & Poor level of connectivity (1), mostly connected (2), fully connected (3). \\
\hline
Segmentation & Segmented 50\% or less (1), 60-80\% (2), more than 80\% of the vessels (3).\\
\hline
Fine Detail & Limited/no small vessel segmentation (1), small gaps in small vessels (2), lots of fine/small vessel detail (3).\\
\hline
False Positives & Lots of (1), limited (2), minimal false positive vessel segmentations (3). \\
\hline
Noise & Lots of noise (1), some/acceptable levels of noise (2), minimal Noise (3).\\
\hline
\end{tabular}
\end{center}
\end{table}

\section{RESULTS}

\subsection{Hysteresis Threshold Range}
\begin{figure}[tbp]
\centerline{\includegraphics[width=\columnwidth]{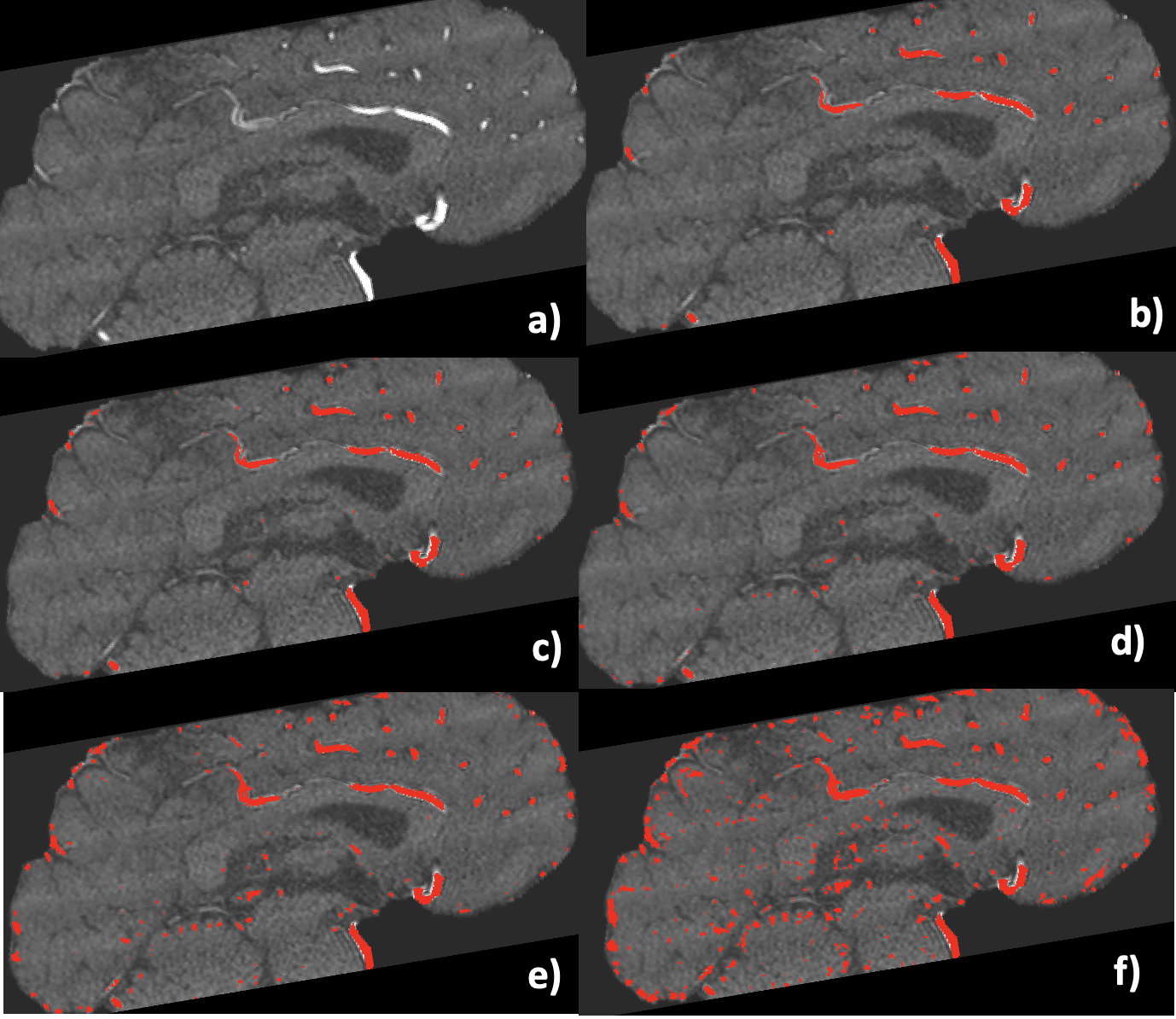}}
\caption{Example of hysteresis thresholds for validation of $\sigma_{u}$ filtered images: a) Original MRA, b) High threshold (H)=69, Low threshold (L)=59, c) H=59, L=49, d) H=49, L=39, e) H=39, L=29, f) H=29, L=19}
\label{fig:thresh}
\end{figure}

The clinician was tasked to select the optimal threshold range for two sets of Hessian filtered images ($\sigma_{l}$ and $\sigma_{u}$) out of five options. Figure~\ref{fig:thresh} is an example of the thresholding study used, with the output segmentation for five threshold ranges provided to the clinician to score alongside the original MRA.
The results from the scoring of the hysteresis thresholds are shown in Table~\ref{tab1} and Table~\ref{tab2}. The spread of clinician preference was one threshold step wider for $\sigma_{u}$ in Table~\ref{tab2}, than for $\sigma_{l}$ in Table~\ref{tab1}. 
 
\begin{table}[tbp]
\vspace*{2.5mm} 
\caption{Percentage of Cases Selected by the Clinician as the Best Hysteresis Threshold Option for ($\sigma_{l}$)}
\begin{center}
\begin{tabular}{|c|c|c|}
\hline
\cline{1-3} 
\textbf{\textit{Lower Threshold \%}}& \textbf{\textit{Upper Threshold \%}}& \textbf{\textit{Clinician Preference}} \\
\hline
\cline{1-3}
77 & 87 & 0\%  \\
\hline
67 & 77 & 17\%  \\
\hline
57 & 67 & 83\%  \\
\hline
47 & 57 & 0\%  \\
\hline
37 & 47 & 0\%  \\
\hline

\end{tabular}
\label{tab1}
\end{center}
\end{table}

\begin{table}[tbp]
\caption{Percentage of Cases Selected by the Clinician as the Best Hysteresis Yhreshold Option for ($\sigma_{u}$)}
\begin{center}
\begin{tabular}{|c|c|c|}
\hline

\cline{1-3} 
\textbf{\textit{Lower Threshold \%}}& \textbf{\textit{Upper Threshold \%}}& \textbf{\textit{Clinician Preference}} \\
\hline
\cline{1-3}
59 & 69 & 0\%  \\
\hline
49 & 59 & 9\%  \\
\hline
39 & 49 & 58\%  \\
\hline
29 & 39 & 33\%  \\
\hline
19 & 29 & 0\%  \\
\hline
\end{tabular}
\label{tab2}
\end{center}
\end{table}

\subsection{Ablation Study}
The results of the ablation study are shown in Table~\ref{tab3}, with the full method achieving the highest quality score. The output without connected component selection scored lowest but had a lower score variance than the remaining alternatives.

\begin{table}[tbp]
\caption{Average Clinician Quality Score for Alternative Methods}
\begin{center}
\begin{tabular}{|c|c|c|}
\hline

\cline{1-3} 
\textbf{\textit{Alternative Method}} & \textbf{\textit{Quality Score, out of 15}} & \textbf{\textit{Score s.d}}\\
\hline
\cline{1-2} 
{Full Method} & 14.2 & 0.9\\
\hline
{Removal of $\sigma _{u}$} & 12.4 & 1.1\\
\hline
{Removal of $\sigma _{l}$} & 12.8 & 0.9\\
\hline
{No Hysteresis Thresholding} & 14.1 & 1.1\\
\hline
{No Connected Component} & 11.0 & 0.2\\
\hline
\end{tabular}
\label{tab3}
\end{center}
\end{table}

\section{DISCUSSION} 

A hysteresis range of 57\% to 67\% of the 99.9th percentile was determined to be the optimal choice for the $\sigma_{l}$ Hessian filtered image, as the clinician concluded that it was the optimal threshold range for the majority (83\%) of the images displayed. A range of 39\% to 49\% of the 99.9th percentile was chosen for the $\sigma_{u}$ filtered images, as it was considered optimal for the majority (58\%) of the images. 

\begin{figure}[tbp]
\centerline{\includegraphics[width=\columnwidth,trim={0 0 0 -6mm}]{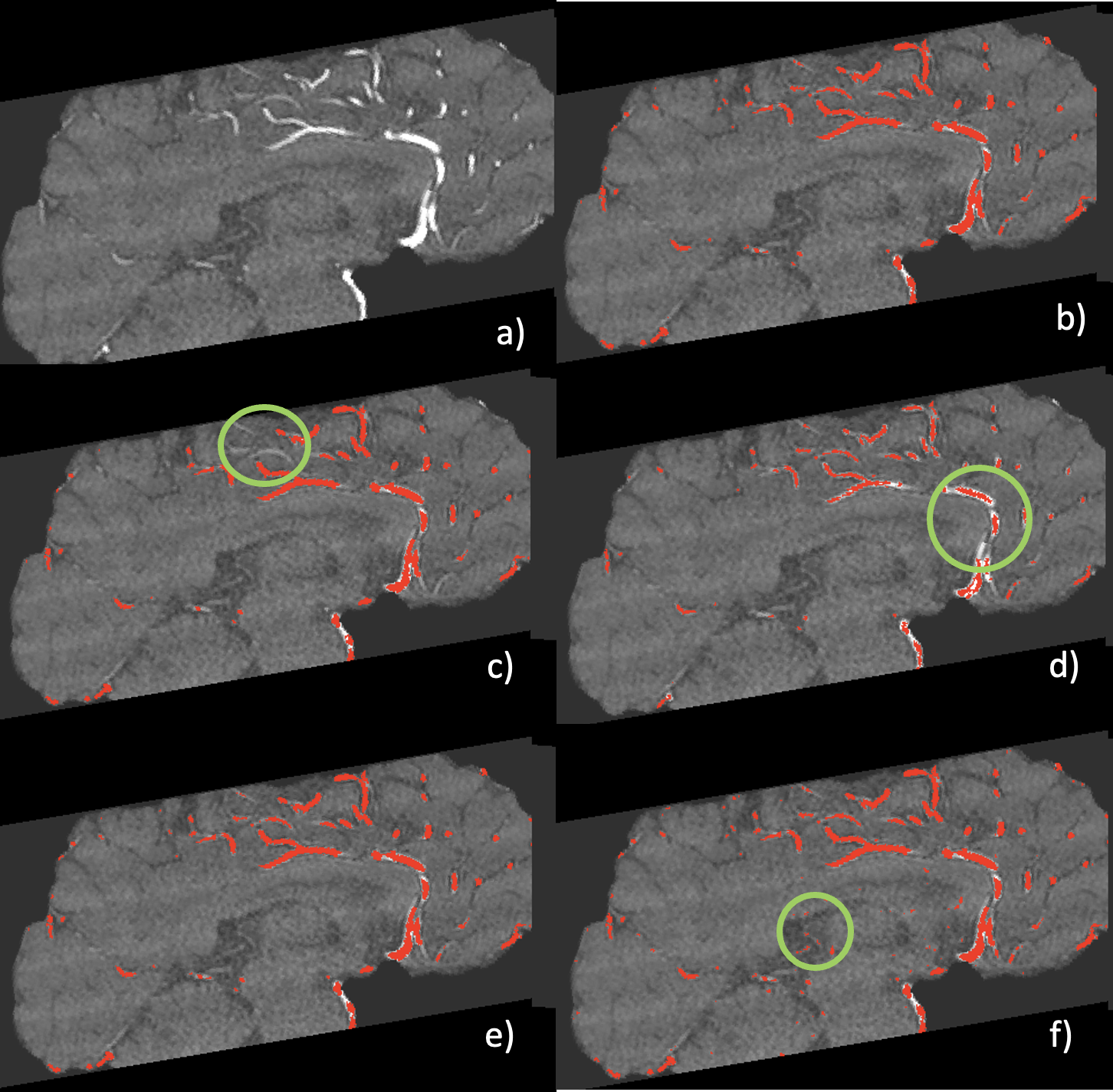}}
\caption{Example slice of alternative method outputs: a) Original MRA, b) Full method c) No $\sigma _{l}$ filter, d) No $\sigma _{u}$ filter, e) No hysteresis thresholding, f) No component correction}
\label{fig:ablation}
\end{figure}

The ablation study confirmed that all steps of the method, in the specific sequence established, were required to optimize the segmentation output. The method produced outputs such as the slices presented in Figure~\ref{fig:ablation} - and the full method achieved the highest quality score, as shown in Table~\ref{tab3}.
Figure~\ref{fig:ablation}~c~\&~d highlighted the necessity for two Hessian filters to be applied to MRA prior to thresholding. The different $\sigma$ values for larger and smaller vessels optimized the thresholding in different ways. The green rings highlighted complete loss of small vessel detail when a filter of $\sigma_{l}$ was omitted, and under-prediction of larger vessels when the $\sigma_{u}$ filter output was removed. The study corroborated the literature~\cite{avadiappan2020fully} that noted the importance of using different sigma values for Hessian filters for vessels of different radii. 
Removal of hysteresis thresholding was shown to have the smallest effect on vessel segmentation, with a score of 14.1 compared to 14.2 for the full method. However, the maximum score was only achieved with the full method, and examples presented to the clinician highlighted the loss of vessel edges when the hysteresis thresholding step was omitted, confirming a similar result in retinal imaging~\cite{chang2008small}. 

Morphological operations (e.g. erosion) were not used due to the thin structure of the vessels being eroded too aggressively. The connected component correction solved this issue by removing clusters of noise without affecting the structural integrity of the remaining vessels. The ablation study showed that component correction was the method's most important step, with the most severe quality loss compared to removing any of the other steps (score went from 14.2 to 11.0). The green ring over the MRA segmentation f) in Figure~\ref{fig:ablation} highlights the advantage of the step, as lots of noise remains on removal of it. This confirms the importance of removing scattered errors, as noted by Lu et al.~\cite{lu2016vessel}.  

Future work will use more participants, from different centers and data sets and fine-tune the parameters. However, although the quality score of 14.2 out of 15 leaves some room for improvement, this may only be achievable with deep learning. Our method provides sufficient ground truth for the training of a deep learning model on both MRA and non-MRA images, and future DL models can surpass this performance by using these segmentations as weak or noisy labels, and generalize these across even larger data sets.

\section{CONCLUSION}
We presented and evaluated a novel method for the automated segmentation of cerebral blood vessels from MRA images. The method and IXI segmentations are available on GitHub, and can be used as weak/noisy labels for training deep learning models to segment MRA and non-MRA images. 

\section*{ACKNOWLEDGMENT}
We would like to thank Minyan Zeng from the Medical School at The University of Adelaide for the blinded labelling. 

\bibliography{ref.bib}
\bibliographystyle{IEEEtran}

\end{document}